# Shock excitation of the knots of Hen 3-1475 [*]

Angels Riera[1,2], Luc Binette[3], and Alejandro C. Raga[4]

[1] Departament de Física i Enginyeria Nuclear, Universitat Politecnica de Catalunya, Av. Víctor Balaguer s/n, E-08800 Vilanova i la Geltrú, Spain
e-mail: angels.riera@upc.edu

[2] Departament d'Astronomia i Meteorologia, Universitat de Barcelona, Av. Diagonal 647, E-08028 Barcelona, Spain

[3] Instituto de Astronomía, UNAM, A. Postal 70-264, 04510 México, México
e-mail: binette@astroscu.unam.mx

[4] Instituto de Ciencias Nucleares, UNAM, A. Postal 70-543, 04510 México, México
e-mail: raga@nucleares.unam.mx

**Abstract.** We present new optical STIS HST spectroscopic observations of the jets of the proto-planetary nebula Hen 3-1475. The excitation conditions of the knots of Hen 3-1475 are derived from the observed optical spectra, confirming that the knots are shock excited. The shocked spectra are qualitatively reproduced by simple "3/2"D bow shock models. We present a set of bow shock models devoted to planetary nebulae, and discuss the effects of the pre-ionization conditions, the bow shock velocity, the bow shock shape and the chemical abundances on the predicted spectra. To explore the reliability of the "3/2"D bow shock models, we also compare the observed spectra of other three proto-planetary nebulae (M 1-92, M 2-56 and CRL 618) to the predicted spectra.

**Key words.** ISM: jets and outflows — planetary nebulae: individual: Hen 3-1475

## 1. Introduction

High spatial resolution images of proto-planetary nebulae (PPNe) and young planetary nebulae (PNe) obtained with the HST, have revealed that a significant fraction show bipolar or mutipolar lobes, highly collimated outflows or jets, and other complex structures (see, e.g., Sahai & Trauger 1998; Garc´ıa-Lario, Riera & Manchado 1999; Trammell & Goodrich 2002). The emission arising from bow shock-like structures, bullets and jets in PPNe and young PNe (i.e. prior to the onset of the photoionization) is likely to be shock excited. Bow-shock like regions with shocked gas have also been observed in more evolved PNe, as in M2-48 (V´azquez et al. 2000) and K 4-47 (Goncalves et al. 2004).

The bow-shock features observed in OH 231.8+4.2 (see, e.g., Bujarrabal et al. 2002), the high-velocity knots, moving outwards with velocities of 70 - 90 km s$^{-1}$ with respect to the central source and observed at both ends of the bipolar lobes of M 1-92 and M 2-56 (Trammell, Dinerstein & Goodrich 1993 (TDG), Trammell & Goodrich 1996, Bujarrabal et al. 1997, Trammell & Goodrich 1998, Castro-Carrizo et al. 2002) are shock-excited features, which are the optical manifestation of the interaction of a jet (or bipolar outflow) with the surrounding material.

Optical jets have been observed in the proto-planetary nebulae CRL 618 and Hen 3-1475. The HST images of CRL 618 have revealed the presence of narrow lobes at different orientations perhaps resulting from multiple ejections at different orientations. Several bow like structures are seen within the body of the lobes of CRL 618, and two bow-shaped structures observed at the tips of the lobes (Trammell & Goodrich 2002). Hen 3-1475 shows the highest outflow velocities ever observed in a PPNe (up to 1200 km s$^{-1}$, Borkowski & Harrington 2001, and Riera 2004). The jets in Hen 3-1475 consist in a S-shaped string of shock-excited knots (Borkowski, Blondin & Harrington 1997; Riera et al. 2002, 2003), whose emission forms in a shock wave propagating at velocities from ∼ 100 to 150 km s$^{-1}$ (Riera et al. 1995, 2003). The inferred shock velocities are apparently inconsistent with the large radial velocities of these knots. These two observational facts become consistent with each other if the knots arose from a variable ejection velocity source (possibly periodic), as was simulated by Vel´azquez, Riera & Raga (2004).

In this paper, we present new results concerning the physical conditions of the jets in Hen 3-1475 obtained from low spectral resolution STIS spectra. The observations are de-scribed in Section 2. From the observed emission line ratios, we calculate the extinction towards Hen 3-1475 (in Section 3) and derive the physical conditions (in Section 4). The analysis of the shock-excited knots of Hen 3-1475 in terms of a '3/2'D bow shock models requires plane-parallel shock models with PNe abundances and the assumption of a bow-shock shape. In section 5, we present the bow-shock models and we discuss the emission line ratios these predict. In subsection 6.1, we compare our bow shock models with the STIS observations of Hen 3-1475. These models are compared in subsection 6.2 with observations of other well-known proto-planetary nebulae (M 1-92, M 2-56, and CRL 618). Finally, in section 7, we present our conclusions.

## 2. Observations

The Space Telescope Imaging Spectrograph data were retrieved from the HST archives. The STIS spectra were taken along the jet axis in Hen 3-1475 on 1999 June 16 and form part of the proposal 7285 (P.I. Harrington). The G430L and G750L gratings were used. The G430L grating was centered on $\lambda_c = 4300$ Å, covering the wavelength region from 2900 to 5700 Å. The G750L grating was centered on $\lambda_c = 7751$ Å, covering the wavelength range from 5236 to 10266 Å. The dispersion was 2.73 and 4.92 Å pixel$^{-1}$ for the G430L and G750L gratings, respectively. The scale along the slit was 0″.05. The 0″.2 width slit was positioned along the jet axis and the spectra were obtained at four parallel positions with a separation offsets of 0″.2 (offsets of −0″.2, 0″.0, +0″.2 and +0″.4 with respect to the central star). The four spectra (which cover NW1, SE1 and SE2) were co-added. In the following, we use the nomenclature of the knots in the NW and SE lobes introduced by Riera et al. (2003).

The long-slit STIS spectra were wavelength and flux calibrated at the Space Telescope Science Institute following the standard HST pipeline calibration. One-dimensional spectra were obtained by summing across each knot. The projected sizes of the knots are: 1″.4 (NW1a, the brightest condensation of knot NW1), 1″.5 (SE1a), 0″.85 (SE1b+SE1c), and 1″.0 (SE2). The reduced spectra are shown in Figures 1 and 2.

Line fluxes were measured by fitting gaussians to the observed line profiles using the IRAF package. The knots of Hen 3-1475 were shown by Riera et al. (2003) present large velocities and wide emission line profiles. At the lower spectral resolution of 4.92 Å (equivalent to ∼ 450 km s$^{-1}$ at Hα) provided by the G750L grating, the [N II] 6548, 6583 Å and Hα 6563 Å emission lines are significantly blended. Despite this caveat, the fitting of double profiles using gaussians is an acceptable procedure since restrictions can be applied to the fit. We fixed both the [N II] 6548/6583 flux ratio to 2.88 and the relative line positions of the three lines [N II] 6548, 6584 Å and Hα. Introspection by eye of the blended profiles clearly reveal that [N II] 6583 line is much brighter than Hα. The deblending procedure adds uncertainty to the [N II] 6548/Hα line ratios, which in our estimate does not exceed 30%.

The resulting emission line ratios are listed in Tables 1 and 2, for the blue and red spectra, respectively.

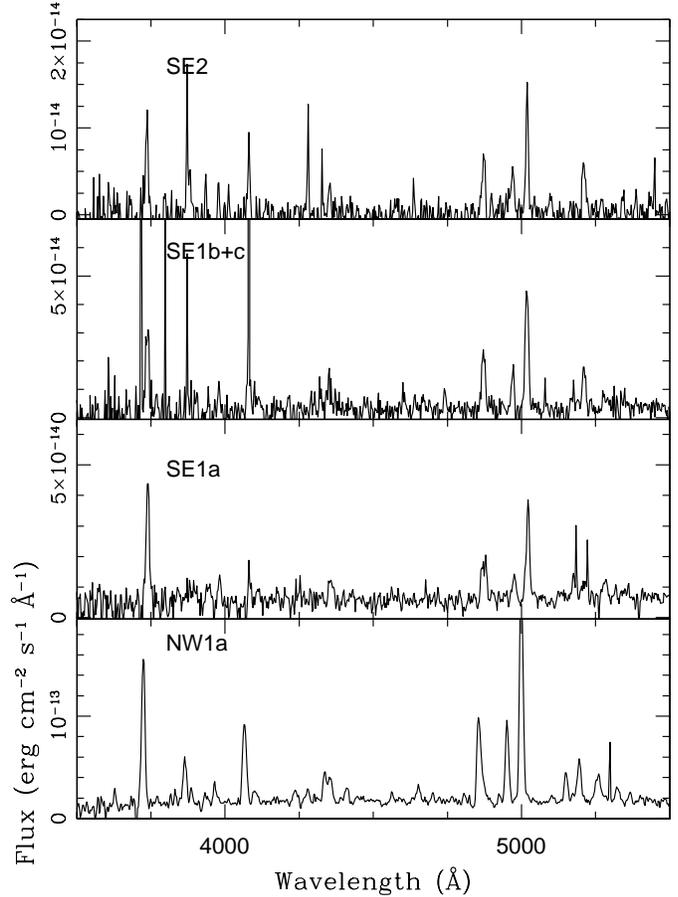

**Fig. 1.** The blue low resolution spectra of the knots NW1a, SE1a, SE1b+c and SE2 of Hen 3-1475 (from 3500 to 5500 Å). The forbidden emission lines of [O II] 3727 Å, [O III] 4960, 5007 Å and [N I] 5200 Å are clearly present in these spectra.

## 3. Extinction

Tables 1 and 2 show the observed blue and red spectra of the observed knots with respect to Hβ = 1 and Hα = 1, respectively. The line flux errors listed in these tables were deduced from gaussian line profile fitting and did not include uncertainties in the flux calibration, and are therefore likely to be underestimated.

We need to determine the E (B - V) value in order to deredden the observed spectra. There are different methods by which this extinction can be derived. In this case, the value of E (B - V) has been determined by assuming the appropriate intrinsec value for the HI Balmer decrements. We have assumed that the intrinsic Hγ/Hβ ratio corresponds to the recombination cascade value. It is expected that the Hγ/Hβ ratio does not deviate appreciably from the recombination value for shock velocities in the range 100–250 km s$^{-1}$ (Hartigan, Raymond and Hartmann 1987, HRH), that is, the same range expected for the shock emission in Hen 3-1475 (Riera et al. 1995, 2002, 2003).

The observed Hγ/Hβ ratio of 0.27 and 0.32 in the spectra of NW1a and SE1a respectively, imply E(B-V) values of 1.2±0.3 and 0.85±0.35, which correspond to visual extinctions of A$_v$ ∼ 3.7±0.9 and 2.6±1.0 mag for the NW1a and SE1a knots. These values are compatible with the extinction derived from the $H_\alpha/H_\beta$ Balmer decrement by Riera et al. (1995), who determined $A_v \sim 3.4\pm0.6$ and $2.0\pm0.4$ for the NW2 and SE2 knots, respectively.

**Table 1.** Observed Blue Spectra relative to $H_\beta = 1$

| Emission Line Identification | NW1a | SE1a | SE1b+c | SE2 |
|---|---|---|---|---|
| [$OII$] 3727 | 1.5±0.03 | 1.3±0.4 | 1.25±0.30 | 1.68±0.23 |
| $H_\gamma$ | 0.27±0.02 | 0.32: | – | – |
| [$OIII$] 4363 | <0.336 | – | – | – |
| $H_\beta$ | 1.0 | 1.0 | 1.0 | 1.0 |
| [$OIII$] 5007 | 2.20±0.02 | 1.2±0.1 | 1.50±0.15 | 1.44±0.12 |
| [$NI$] 5200 | 0.42±0.05 | – | 0.65±0.05 | 0.90±0.05 |

**Table 2.** Observed Red spectra relative to $H_\alpha = 1$

| Emission Line Identification | NW1a | SE1a | SE1b+c | SE2 |
|---|---|---|---|---|
| [$NII$] 5755 | 0.17±0.02 | 0.14±0.04 | 0.068±0.006 | 0.134±0.002 |
| He I 5876 | – | 0.093: | 0.092: | – |
| [$OI$] 6300 | 0.67±0.06 | 0.285±0.030 | 0.48±0.08 | 0.68±0.01 |
| $H_\alpha$ | 1.0 | 1.0 | 1.0 | 1.0 |
| [$NII$] 6583 | 3.7±0.3 | 2.40±0.20 | 2.3±0.4 | 3.2±0.2 |
| [$SII$] 6717+6731 | 0.68±0.04 | 0.66±0.08 | 0.86±0.04 | 0.76±0.03 |
| He I 7065 | 0.023: | – | – | – |
| [$CaII$] 7290 | 0.14±0.02 | 0.20±0.03 | 0.18±0.02 | – |
| [$OII$] 7320-30 | 0.76±0.07 | 0.39±0.03 | 0.41±0.07 | 0.61±0.01 |
| F($H_\alpha$) $10^{12}$ | 6.4 | 2.0 | 3.0 | 1.5 |

Finally, the observed blue and red dereddened ratios are shown in Tables 3 and 4. The adopted E (B-V) values are 1.2 and 0.85 for the NW knots and SE knots, respectively. Due to the low S/N of the blue spectra, we have not considered extinction variations along the SE jet. The uncertainties in the dereddened emission line ratios do not include the uncertainties resulting from the determination of E (B-V).

## 4. Observed emission lines and excitation conditions

The strongest lines in the visible spectra of Hen 3-1475 are, apart from the Balmer lines of hydrogen, [OI] 6300, 6363 Å, [O II] 3727, 7320-30 Å, [O III] 4959, 5007 Å, [N II] 5200 Å, [N II] 5755, 6548, 6583 Å, and [S II] 6716, 6730 Å.

Electron densities are computed from the [S II] 6717/6731 ratios following the standard procedure. The [S II] 6717, 6731 Å emission lines, which overlap at the spectral resolution given by the G750L grating, were measured instead from the STIS spectra obtained with the G750M grating (Borkowski & Harrington 2001, Riera et al. 2003). The results are listed in Table 5. For the subcondensations of knot SE1 the electron densities range from 1300 to 1900 cm$^{-3}$, while a higher density is derived for knot NW1a ($\sim 3000$ cm$^{-3}$). Since the compression factor across a strong shock is four, we can estimate the preshock densities from the electron densities quoted above. We therefore infer that the preshock densities lie in the interval $\sim 300$ to 500 cm$^{-3}$ (slightly higher for knot NW1a). In the following, we will adopt a value of 400 cm$^{-3}$ for the preshock gas (see Section 5). The highest electron density is found at knot

SE2 ($\sim 5800$ cm$^{-3}$), which is larger than the electron density of $\sim 3000$ cm$^{-3}$ detected from ground-based observations of the intermediate knots (Riera et al. 1995).

The electron temperatures have been derived from the [N II] (6548+6583)/5755 emission line ratios. The dereddened [N II] 6548+6583/5755 ratios vary from 15.4 at SE1a to 31.0 at SE1b+c. From these emission line ratios, we inferred T ([NII]) values of 25200 K (NW1a), 26900 K (SE1a), 16500 K (SE1b+c) , and 20100 K (SE2). These values are compatible with electron temperatures previously computed from ground based observations of SE2 and NW2 (Riera et al 1995) despite the uncertainties resulting from the gaussian fitting procedure of the blended $H_\alpha$ and [N II] lines.

## 5. Bow-shock models

The emission line spectra and line profiles together with the geometry of the knots of Hen 3-1475 and their high velocities suggest the possibility of interpreting these knots as bow shocks.

The double-peaked emission line profiles and the extraordinarily large line widths observed in the knots of Hen 3-1475 (Borkowski & Harrington 2001, Riera et al. 2003, Riera 2004) resemble the properties observed in several HH objects, which have been successfully explained in terms of bow shock models (see, e.g., Raga & Böhm 1985, 1986, HRH). A bow shock has the advantage of producing wide emission line profiles in a small volume.

Bow shocks can be formed by the interaction of clumps of gas with the streaming ambient medium during discrete events of ejection ('interstellar bullet' model; Norman & Silk 1979) or by the interaction of an outflow with a clumpy medium ('stationary shocked cloudlet'; Schwartz 1978). The structure

**Table 3.** Dereddened blue emission line intensities relative to H$_\beta$ = 1

| Emission Line Identification | NW1a | SE1a | SE1b+c | SE2 |
|---|---|---|---|---|
| [OII] 3727 | 4.36±0.09 | 2.8±0.8 | 3.2±0.8 | 3.6±0.5 |
| [OIII] 4363 | <0.81 | – | – | – |
| H$_\beta$ | 1.0 | 1.0 | 1.0 | 1.0 |
| [OIII] 5007 | 1.92±0.02 | 1.09±0.09 | 1.36±0.14 | 1.3±0.1 |
| [NI] 5200 | 0.30±0.04 | – | 0.52±0.03 | 0.73±0.04 |

**Table 4.** Dereddened red emission line intensities relative to H$_\alpha$ = 1

| Emission Line Identification | NW1a | SE1a | SE1b+c | SE2 |
|---|---|---|---|---|
| [N II] 5755 | 0.29±0.03 | 0.21±0.06 | 0.10±0.01 | 0.19±0.02 |
| He I 5876 | – | 0.15 | 0.15 | – |
| [O I] 6300 | 0.78±0.07 | 0.32±0.03 | 0.53±0.09 | 0.76±0.02 |
| H$_\alpha$ | 1.0 | 1.0 | 1.0 | 1.0 |
| [N II] 6583 | 3.6±0.3 | 2.4±0.2 | 2.3±0.4 | 3.2±0.2 |
| [S II] 6717+6731 | 0.62±0.04 | 0.62±0.08 | 0.81±0.04 | 0.72±0.03 |
| He I 7065 | 0.017 | – | – | – |
| [CaII] 7290 | 0.09±0.01 | 0.13±0.02 | 0.12±0.01 | – |
| [O II] 7320-30 | 0.53±0.05 | 0.30±0.02 | 0.32±0.05 | 0.48±0.01 |

**Table 5.** High spectral resolution: Observed emission line ratios

| Observations | Knot + component | [S II] 6717/6731 | n$_e$ ([SII]) cm$^{-3}$ |
|---|---|---|---|
| STIS HST | NW1a | 0.64 | 2960 |
| STIS HST | SE1a | 0.79 | 1260 |
| STIS HST | SE1b+c | 0.71 | 1880 |
| STIS HST | SE2 | 0.56 | 5820 |

–morphology and overall kinematics– of Hen 3-1475 can be reasonably fitted by a time-dependent jet model with a slow precession and with an ejection velocity history composed of a sinusoidal mode superimposed on a linear ramp, as shown by the numerical simulations of Vel´azquez, Riera & Raga (2004). In this scenario, the jet fragments into clumps that behave like 'interstellar bullets', which can, therefore, be modeled as bow shocks.

The models widely applied to HH objects are the '3/2-D' bow shock models, which have been successfully used to model the emission line ratios and line profiles in HH objects (e.g., Raga & Böhm 1985, 1986, HRH). In these models, one assumes a surface that approximately reproduces the shape of the bow shock. As the characteristics of the flow that produces the bow shock (e. g., the cross section of the jet, see Raga, Cant´o and Cabrit 1998) are not known, a priori a wide range of bow shock shapes are in principle possible. Therefore, it is correct to choose the bow shock shape that best fits the observed line ratios. A schematic diagram of the bow shock geometry is shown in Fig. 3. Following the '3/2-D' bow shock prescription, we computed a set of models devoted to proto-planetary nebulae. We model the emission by a set of planar shocks with shock velocities given by the component of the velocity normal at each point along the bow shock. To calculate the line emission, the bow shock was divided into 200 annuli. The emission from each annulus is taken to be that of the planar shock with the corresponding normal shock velocity weighted by the area of the annulus. We model the emission by interpolating (linearly) among a series of planar shock models.

We assume that the shock wings extend to large distances from the apex of the bow shock, which is moving at a velocity $V_{bs}$ with respect to the pre-shock medium. The pre-shock gas is assumed to have a constant density. We have considered two sets of chemical abundances, which correspond to the mean Type I and Type II PNe abundances, respectively (from Kingsburgh & Barlow 1994).

### 5.1. The bow-shock shape

We consider the following functional form to describe the bow shock surface:

$$\frac{z}{a} = \left(\frac{r}{a}\right)^p \tag{1}$$

where $z$ is measured along the symmetry axis and $r$ is the cylindrical radius. The constants $a$ and $p$ are free parameters of the model and were described by Beck et al. (2004). The parameter $p$ is a form factor that determines the shape while the constant $a$ determines the size of the bow shock. Figure 4 illustrates this shape for $p = 2$, 3 and 4. Recently, Schultz, Burton and Brand (2005) have studied the effect of the bow shape on the H$_2$ emission line profiles by comparing the predictions of the parabolic and cubic bow shapes, and the shapes adopted by Raga and Böhm (1985) and HRH. They concluded that the cubic, the Raga and Böhm and HRH bow shapes give similar line profiles.

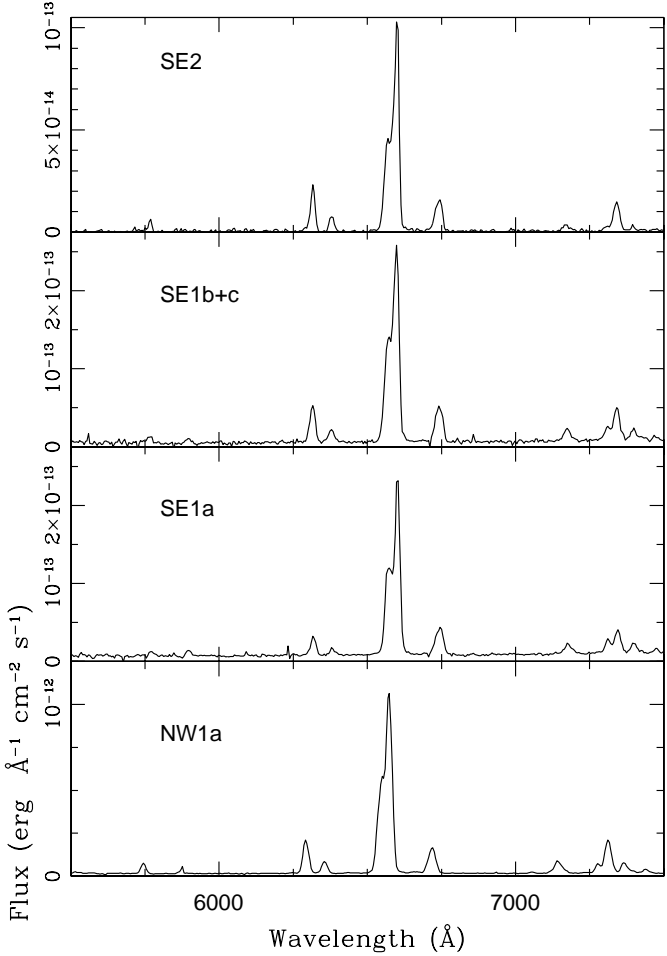

**Fig. 2.** The red low resolution spectra of the knots NW1a, SE1a, SE1b+c and SE2 of Hen 3-1475 (covering from 5500 to 7500 Å). Forbidden emission lines of [N II] 5755, 6548, 6583 Å, [O I] 6300, 6363 Å, [S II] 6716, 6730 Å, and [O II] 7320-30 Å are present in these spectra.

### 5.2. Plane-parallel steady shock models

The planar shock models that we used to predict the bow shock emission line ratios were obtained with the photoionization-shock code MAPPINGS Ic (Binette, Dopita & Tuohy 1985; Ferruit et al. 1997). The variation of the perpendicular velocity across the bow shock requires a grid of models closely spaced in shock velocity. Hence, planar shock models were obtained for shock velocities from 20 to 300 km s$^{-1}$ at intervals of 10 km s$^{-1}$. We have adopted a pre-shock density of 400 cm$^{-3}$. The pre-shock magnetic field was taken to be quite small (B =0.1 $\mu$G).

We have adopted either the mean Type I or mean Type II PN abundance ratios, both from Kingsburgh & Barlow (1994). As the abundances of Fe, Mg, Si and Ca were not measured by Kingsburgh & Barlow (1994), these were set to their cosmic values. Notice that both sets of abundances differ by their He/H and N/H abundance ratios, which are 1.15 and 3.8 times larger for the Type I PNe, respectively.

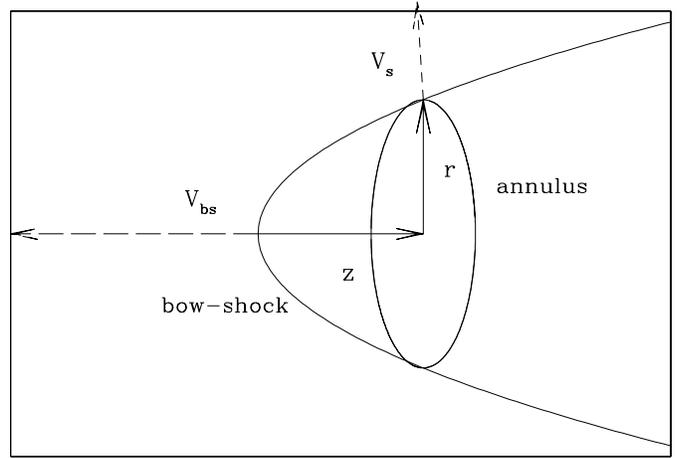

**Fig. 3.** A schematic diagram of the bow shock geometry. The bow shock moves at a speed V$_{bs}$ in the negative z-direction. V$_s$ is the component of the velocity normal to the bow shock.

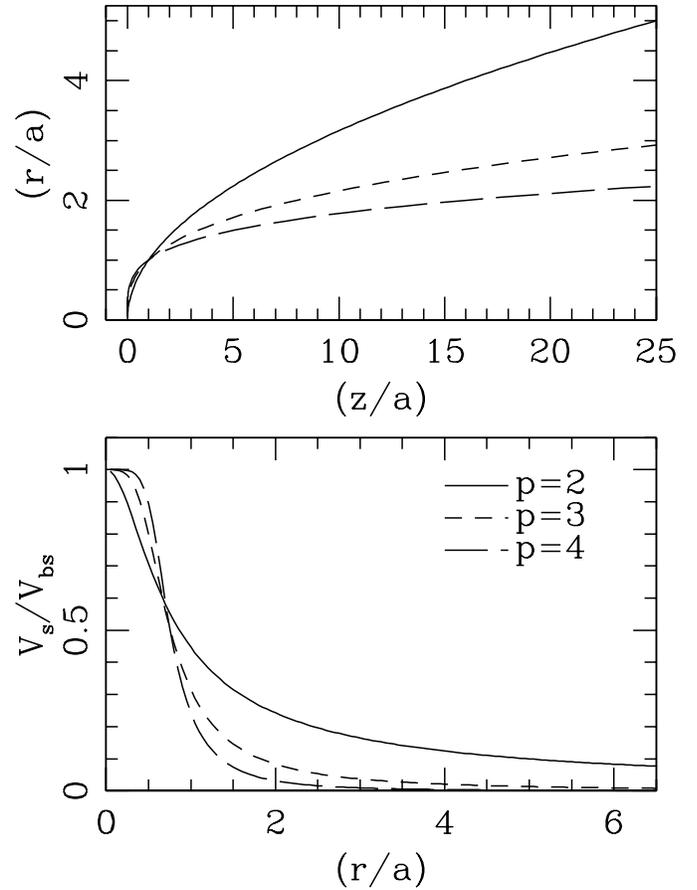

**Fig. 4.** Top panel: the bow shock shape for $p = 2$, 3 and 4. Bottom panel: the shock velocity (i.e. the component of the velocity normal to the bow shock surface) as a function of $r$.

The predicted spectrum from a shock front strongly depends on the ionization state of the pre-shock gas. Two pre-ionization conditions were considered.

First, the pre-shock ionization fractions for each value of $V_s$ (i.e. the effective shock velocity on a given annullus, see figure 3) was taken to be the self-consistent value determined by the ionizing flux emitted by the shock at that velocity (plane-parallel 'local equilibrium preionization" condition, as described by Shull & McKee 1979, and HRH). In this scenario the ionization fractions change from high values at the apex of the bow-shock to lower values at the wings. However, this situation does not apply to non-planar configurations where the pre-shock gas in the wings of the bow-shock is exposed to the ionizing radiation from parts of the bow shock that have different effective shock velocities.

The problem of the preionization in non-plane-parallel flow configurations did not received much attention in the past, although there has been a few attempts to determine the preionization of the regions upstream of a bow shock. For instance, Raymond, Hartigan & Hartmann (1988) computed, as a function of distance to the axis, an approximate radiative transfer for preionized regions upstream of the bow shock. They found that the preshock gas to be less ionized near the bow shock apex and more ionized in the wings than if would be in the plane-parallel 'local equilibrium preionization" case.

A more realistic treatment of the radiative transfer was developed by Raga et al. (1999), who applied a 3D radiative gas-dynamic code with diffuse ionizing radiation, to the determination of the preionization in a bow shock. The photons produced in the post shock region that are diffusing upstream produce a radiative precursor, which is partially ionized and quite warm. They found that –for a bow shock velocity of 200 km s$^{-1}$– the H$\alpha$ emission of the radiative precursor is much fainter than the emission from the recombination region and therefore can be neglected. Numerical simulations that include heavy atomic elements and explore the effect of varying the bow shock velocity parameter are urgently needed in order to determine whether the radiative precursor might appreciably emit forbidden lines.

Having in mind the problem of the preionization of non-planar shocks and in order to explore the dependence of the bow shock spectrum on preionization conditions, we computed another set of models in which the pre-shocked gas is assumed to be uniformly ionized (i.e. an ionization that does not vary with distance to the apex). More specifically, we set the ionization fractions of H$^+$ and He$^{++}$ to 1. We refer to this set of models as 'fully" preionized models.

### 5.3. Bow shock results: predicted spectra as a function of the preionization conditions, bow-shock velocity and bow shape

In this section, we present a number of emission line ratio diagnostics. In order to characterize the optical shock spectra assuming different pre-ionizations conditions, we have adopted the Type I PNe abundances (from Kingsburgh & Barlow 1994) and two preionization conditions ('local equilibrium" and 'full preionization"). The adopted bow-shock velocities range from 50 to 300 km s$^{-1}$. The bow shape is determined by the parameter $p$, for which values of 2, 3 and 4 were adopted (see Fig.4).

The results are shown in Fig. 5, where we present several emission line ratios as a function of [N II] 6583/H$\alpha$. We have choosen the [N II] 6583/H$\alpha$ emission line ratio as the abscissa because the [N II] 6583 Å line is a strong line in most PPN (see below). The predicted spectrum strongly depends on the preionization conditions. In the following, we point out the main effects of the preionization conditions, the bow shock velocity and its shape on the predicted spectra.

#### 5.3.1. Preionization conditions

We should note that for the 'equilibrium preionization" models, the [N II] 6583/ H$\alpha$ emission line ratio shows a large range of values (from 0.20 up to 2), while the 'full preionization" models are characterized by strong [N II] 6583/ H$\alpha$ ratios, with values between $\sim$ 2 and 3.

The [N I] 5200/H$\beta$ vs. [N II] 6583/ H$\alpha$ diagnostic plot in Fig. 5 shows a clear separation between 'local equilibrium preionization" and 'full preionization" models. As expected, the 'local equilibrium preionization" set of models predict larger [N I] 5200/H$\beta$ ratios than the 'fully preionized" models, in which NI occurs only in the downstream recombining region of the shock and the predicted [N I] 5200/H$\beta$ turns out then much smaller ($\leq 0.5$).

On the [O II] 3727/H$\beta$ vs. [N II] 6583/H$\alpha$ diagram, different classes of shocks (i.e. different preionization conditions) show a clear separation. The 'local equilibrium preionized" models are found in the bottom left-side region of the plot (where [O II] 3727/H$\beta \leq 1$, and [N II] 6583/H$\alpha \leq 2$). Qualitatively, the main effect of considering that the preshock gas is fully ionized is to increase the [O II] 3727/H$\beta$ and [N II] 6583/H$\alpha$ ratios (see Fig.5).

In the diagnostic diagrams of high excitation emission lines (i.e. [O III] 5007 Å, [Ne III] 3869 Å, and He II 4686 Å) with respect H$\beta$ vs. [N II] 6583/H$\alpha$, the behaviour of the different types of bow shock models is again distinct. The 'fully preionized" models fall on the upper region of the diagnostic plots for these high excitation lines as well (i.e.'fully pre-ionized" models predict the largest values for the high excitation line ratios).

Shock models with different preionization conditions predict similar values for several low-excitation emission line ratios (i.e., [O I] 6300/H$\alpha$, [S II] (6717+6731)=(6725)/H$\alpha$, [Ca II] 7291/H$\alpha$).

#### 5.3.2. Bow shock velocity

For the 'equilibrium preionization" models, the [N II] 6583/ H$\alpha$ emission line ratio increases with increasing bow-shock velocity. The 'full preionization" models are characterized by strong [N II] 6583/ H$\alpha$ ratios which are approximately independent of the bow-shock velocity. For the 'full preionization" models, the [O II] 3727/H$\beta$ emission line ratio decreases as the bow shock velocity increases.

The 'local equilibrium preionization" set of models predict large [N I] 5200/H$\beta$ ratios at low bow shock velocities. In the 'fully preionized" models, the predicted [N I] 5200/H$\beta$ is more or less independent of the value of the bow-shock velocity. As expected, the "local equilibrium preionization" set of models predict large [N I] 5200/Hβ ratios at low bow shock velocities, because NI is an abundant specie in the postshock region. As the bow shock velocity increases, the incoming neutral N becomes more collisionally ionized and, consequently, the [N I] 5200/Hβ decreases as the [N II] 6583/ Hα ratio increases.

For "local equilibrium preionization", the high excitation lines (i.e. [O III] 5007 Å, [Ne III] 3869 Å, and He II 4686 Å) are negligible for $V_{bs} \leq 75$ km s$^{-1}$. For bow-shock velocities above this threshold, these emission line ratios show a rather complex behaviour, depending on the values of the bow shock velocity and on the bow shock shape. The intensity of the high excitation lines with respect to Hβ could either increase (for $p$ = 2) or decrease ($p$ = 4) with increasing bow shock velocity.

The [S II] 6725/Hα emission line ratio is approximately independent of the bow-shock velocity for all shock models considered here. The [O I] 6300/Hα and [Ca II] 7291/Hα emission line ratios increase with bow shock velocity (for bow-shock velocities above 100 km s$^{-1}$).

### 5.3.3. Bow shock shape

We should note that for the "equilibrium preionization" models, the [N II] 6583/ Hα emission line ratio depends on the bow-shock shape, reaching values larger by a factor of 3 for $p$ = 4 with respect to the results for $p$ = 2.

The [S II] 6725/Hα emission line ratio is approximately independent of the bow shock shape $p$ for all shock models considered here.

Some of the emission line ratios are strongly dependent on the shape of the bow shock. Several low excitation emission line ratios (e.g. [N I] 5200/Hβ) become larger with $p$ = 2. However, for $p$=2 the predicted [O I] 6300/Hα and [Ca II] 7921/Hα emission line ratios are significantly lower than for the bow shock models with $p$ = 3 or 4. Emission line ratios of the high excitation lines (i.e. [O III] 5007 Å, He I 5876 Å, He II 4686 Å, and [Ne III] 3869 Å) with respect to the HI Balmer emission lines, increase with increasing $p$ values (i.e., as the bow shape becomes blunter).

## 6. Discussion

### 6.1. The shock-excited knots of Hen 3-1475

The optical spectra of the Hen 3-1475 knots have been found to be the result of shock excitation. Following an earlier comparison of the ground-based spectra with those from plane-parallel shock models, Riera et al. (1995, 2003) concluded that the emission forms in a shock wave that is propagating at ∼ 150 km s$^{-1}$ through a nitrogen enriched medium.

In this work, we have presented the results of bow shock models assuming mean Type I PNe chemical abundances. We have explored the influence of the bow-shock velocity, the bow shape and the preionization conditions on the predicted spectra. The dereddened emission line ratios of several knots of Hen 3-1475 (obtained with STIS, listed in Tables 3 and 4) are represented in the plots of Fig. 5.

The bottom three plots of Fig. 5 show the [O I]/Hα, [O II]/Hβ and [O III]/Hβ (i.e. the three observable ionization stages of oxygen) emission line ratios as a function of [N II]/Hα. In the three plots, the Hen 3-1475 knots fall within the region of the "fully preionized" models. The [O III]/Hβ emission line ratio of NW1a is remarkably large, falling well above the values observed in the other knots and predicted by all models.

The [O II] 3727/Hβ emission line ratios of the knots of Hen 3-1475 are only partially reproduced by the "fully preionized" models, as the observed values remain larger than the predicted values. The best fit corresponds to the $p$ = 2 bow shocks (for a large range in bow-shock velocities).

High-velocity bow shock models with large $p$ values are found to reproduce better the [O I]/Hα emission line ratios.

In summary, the oxygen plots show that there remain significant discrepancies of the models with at least some of the data points. We believe that our 3/2-D models probably fail to include all of the geometrical complexities of the knots. It is possible that before being shocked, these knots already contained substructure and density inhomogeneities. Given the non-linearity of reprocessing of the kinetical energy into different emission lines, we expect that the presence of substructures would affect the integrated postshock spectrum. Eventhough our models are pseudo 3/2-D, they still remain an idealized representation of the knot internal geometry.

The observed [N I] 5200/Hβ vs. [N II] 6583/Hα ratios are clearly in agreement with the predictions of the "fully preionized" models. Therefore, our initial assumption about the N/H relative abundance is consistent with the position of the line ratios of the Hen 3-1475 knots in the latter diagnostic plot. We should note that the increase (decrease) in nitrogen abundance would encompass all points in this diagram to regions with higher (lower) [N I]/Hβ and [N II]/Hα, and clearly beyond the location of the knots of Hen 3-1475 in this particular diagram.

The predicted [N II] 6583/5755 emission line ratios, in general, lie above the observed values. The electron temperatures inferred for the theoretical models are predicted to be lower than the values inferred from the observations of Hen 3-1475. The [N II] 6583/5755 emission line ratios of the SE2 and SE1b+c knots are well reproduced by the bow-shock models.

The [S II]/Hα emission line ratios observed in the knots of Hen 3-1475 are ∼ 3 times larger than all the predicted values. It is important to note that shock models so far have consistently been unable to reproduce the [S II]/Hα emission line ratios observed in HH objects (Raga, Böhm & Cantó 1996).

The simple bow shock models presented here with bow-shock velocities ranging from ∼ 150 to 200 km s$^{-1}$, with type I PNe abundances and assuming a full ionization of the pre-shocked gas can roughly reproduce the emission line ratios of several knots of Hen 3-1475 within the observational uncertainties and the modelling approximations. We should note that for "fully preionized" models, several emission line ratios are independent of the bow shock velocity.

The predicted [S II] 6725/Hα ratios remain too low while the [O I] 6300/Hα line ratios appear to require a higher bow shock velocity (of the order of 300 km s$^{-1}$) than the other emission lines.

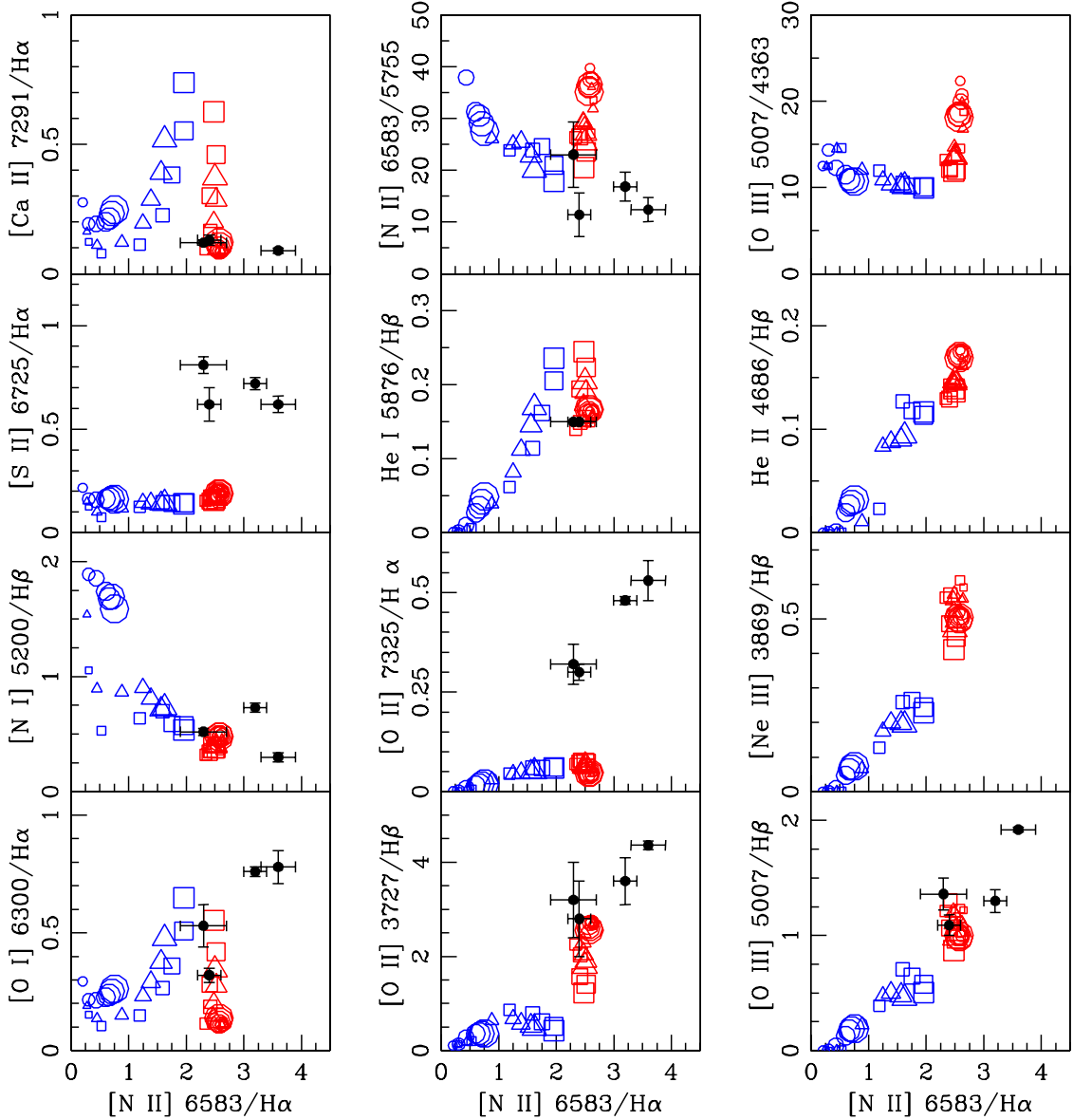

**Fig. 5.** Emission line ratios as a function of [N II] 6583/Hα as predicted from bow shock models, assuming mean Type-I PNe abundances. The "equilibrium preionization" models are shown in blue, and the "full preionization" models are shown as red symbols. The size of the symbol increases with the bow-shock velocity. The adopted values of the bow-shock velocitiy are: $V_{bs}$ = 50, 75, 100, 150, 200, 250, and 300 km s$^{-1}$. Values of the $p$ parameter (see the text) are $p = 2$ (circles), $p = 3$ (triangles), and $p = 4$ (squares). The observed emission lines ratios of the knots of Hen 3-1475 are also represented in these plots as black dots.

The brightest and highest excitation knot (NW1a) is not well reproduced by the bow shock models presented here. The [O III]/Hβ ratio predicted by the model is about 60% too low. The predicted [O I] 6300/Hα and [O II] 3727/Hβ ratios are about 50% lower than the values arising from NW1a. NW1a is the only knot that has been detected in X-ray emission (Sahai et al. 2003). The X-ray spectrum and luminosity of NW1a, assuming that the emission is due to shocks, implies a shock velocity of ∼ 400 to 500 km s$^{-1}$ (Guerrero, Chu & Gruendl 2004; Sahai et al. 2003). At this bow shock velocity, the contribution of the ratiative precursor might be detectable. Observational properties of radiative precursors of plane-parallel shocks have been presented by Dopita & Sutherland (1995, 1996), who found that the main effect of adding a precursor to the observed spectra arising from a shock is the increase of the relative intensity of [O III], which may contribute to reproduce the spectrum of NW1a.

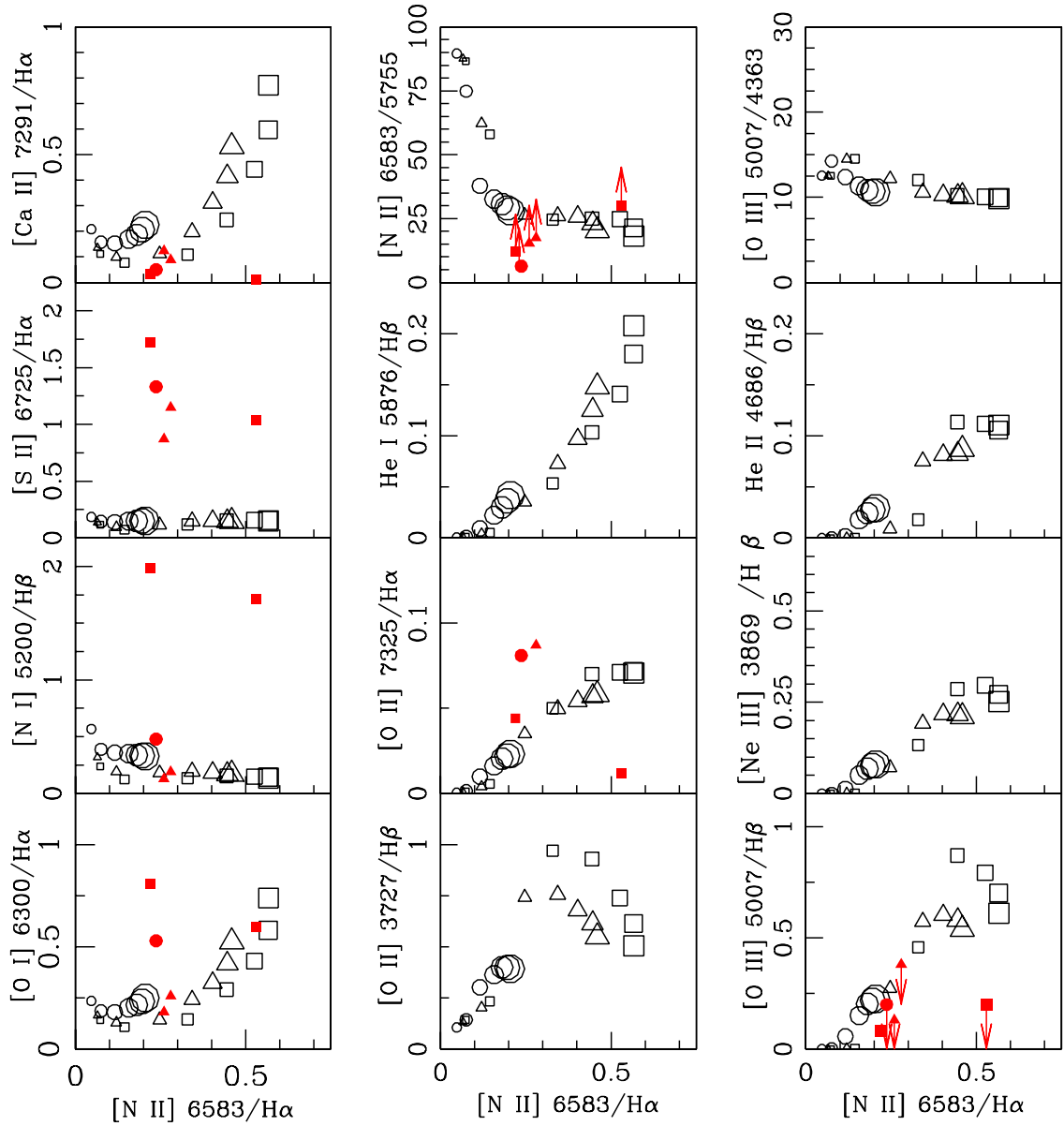

**Fig. 6.** Emission line ratios as a function of [N II] 6583/Hα predicted from bow shock models for mean Type II PNe abundances and "equilibrium preionization" conditions. The observed emisison line ratios of M 1-92 (red triangles), M 2-56 (red circles) and CRL 618 (red squares) are also presented in these plots.

## 6.2. Other Proto-planetary Nebulae

Several proto-planetary nebulae display emission lines likely to be shock-excited. The emission lines arising from the shocked gas of these PPNe were previously analyzed comparing the observed spectra with the shock models devoted to HH objects (see, e.g., TDG; S´anchez Contreras, Sahai & Gil de Paz 2002).

In the following, we compare the modelled emission line spectra to the observed spectra of three PPNe (CRL 618, M1-92 and M2-56) to determine the chemical abundances and the bow-shock velocities of these three objects. A comparison of the predicted spectra with observations would provide an indication of the reliability of our '3/2D' bow-shock models.

The optical spectra of the lobes of these PPNe are composed of a scattered stellar emission plus emission lines. To separate emission originated within the lobes from line emission reflected from the central regions, we use the results of TGD who obtained the unscattered emission of the lobes of the bipolar PPN CRL 618, M 1-92, and M 2-56. Unfortunately TGD obtained one spectra for each lobe by spatially summing across each lobe. For all three nebulae the bow shaped structures/compact knots are the brightest features of these nebulae. Because of this we assume that the spectra obtained by TGD mostly correspond to these shocked features that morphologically resemble bow shock structures. We have presented the emission line ratios of CRL 618, M 1-92, M 2-56 in the plots of Figs. 6 and 7. The [N II] 6583/5755 emission line ratios of the three nebulae are lower limits, and the [O III]/Hβ ratios of the three nebulae are upper limits.

### 6.2.1. The bipolar PPN M 1-92

The narrow band HST optical images of M 1-92 show the presence of compact knots -which are prominent in [O I] and [S II], and faint in [O III]- placed inside the empty cavities detected in CO emission (Bujarrabal et al. 1998, Trammell & Goodrich 1996). The interpretation of these optical images is still controversial. Trammell & Goodrich (1996) interpreted these emitting knots as the points where the collimated outflows impact the bipolar lobes, while Bujarrabal et al. (1998) identified the compact knots with the Mach disk expected to form in the jet as it impacts on the surrounding medium.

Several authors proposed that the optical emission lines observed in M 1-92 correspond to shocks of different shock velocities. The range of shock velocities goes from about 60/100 km s$^{-1}$ (TDG) to about 200/300 km s$^{-1}$ (Solf 1994, Bujarrabal et al. 1998).

The dereddened unscattered emission line ratios of the SE and NW lobes of M 1-92 (taken from TDG) are included in Fig. 6, where we plot the emission line ratios predicted for 'local equilibrium preionization" bow shock models. The chemical abundances adopted correspond to mean Type II PNe abundances.

The observed [N I] 5200/Hβ and [N II] 6583/Hα are in good agreement with intermediate excitation bow shocks (for $p = 3, 4$) (see the corresponding plot in Fig. 6), confirming that M 1-92 will evolve into a Type II PN (as previously stated by TDG).

The M 1-92 points placed in the [O I]/Hα, [Ca II]/Hα and [O III]/Hβ vs. [N II] /Hα plots suggest a bow shock velocity from 100 to 150 km s$^{-1}$ for the SE lobe, and ≤ 100 km s$^{-1}$ for the NW lobe. Once again, the observed [S II] 6725/Hα and [N II] 6583/5755 ratios are above the predicted values.

We conclude that the optical spectra of M 1-92 may arise in a bow shock with velocities from 100 to 150 km s$^{-1}$. The chemical abundances of the shocked gas are consistent with the adopted abundances (i.e. mean Type II PNe abundances).

### 6.2.2. The bipolar PPN M 2-56

The HST images of M 2-56 revealed the presence of several knots in the lobes (Trammell & Goodrich 1998). The emission line ratios of the West lobe of M 2-56 were consistent with the ratios predicted by plane-parallel shock models with shock velocities from 40 to 60 km s$^{-1}$ (TDG). TDG suggested that M 2-56 could be a Type I PN, but the evidence was not conclusive. Castro-Carrizo et al. (2002) proposed that M 2-56 could be an intermediate object between a low-mass post-AGB star and a standard PPN.

The dereddened emission line ratios of the West lobe of M 2-56 (taken from TDG) were plotted in the diagnostic plots of Fig. 6. In general, the observed emission line ratios are marginally reproduced by the bow shock models included in this figure. From the [O III]/Hβ vs. [N II]/Hα we could conclude that the emission is formed in a bow shock model with V$_{bs}$ ∼ 75 km s$^{-1}$ (for 'local equilibrium preionization" conditions and mean Type II PNe abundances). However, the predicted [O I] 6300/Hα and [S II] 6725/Hα ratios are too low. The M 2-56 [N I] 5200/Hβ ratio is about a fator of 2 larger than the predicted values, implying that the N/H abundance of M 2-56 may be higher than the adopted value. Although the [N II] 6583/5755 ratio is a lower limit, we note that the observed ratio is much lower than the predicted values (see Fig. 6).

In order to test if M 2-56 could be nitrogen enriched, we have plotted the emission line ratios of this nebula together with the 'local equilibrium preionization" models (adopting the mean Type I PNe abundance) in Fig. 7. The M 2-56 [N II]/Hα ratio implies a low bow-shock velocity (comparing with the 'equilibrium preionization" Type I PNe abundance set of models). However, the [N I] 5200/β is much lower, and the [O II] 7325/Hα and [O III] 5007/Hβ ratios are larger than the values predicted at the low velocities required by the observed [N II]/Hα ratio.

We find a good qualitative agreement between the predictions of the bow-shock models with V$_{bs}$ ∼ 75 km s$^{-1}$, 'equilibrium preionization" and Type II PNe abundances and the observations of M 2-56. The fitting could be improved if the N/H abundance was a free parameter. In this way, we would find that the best fit to the observations of M 2-56 corresponds to a N/H abundance in between the two values adopted above (corresponding to the mean Type II and the mean Type I PNe abundances as determined by Kingsburgh & Barlow (1994)).

### 6.2.3. The bipolar PPN CRL 618

Ground-based optical images of CRL 618 show two lobes of emission separated by a dark lane. Its optical HST images revealed the presence of narrow lobes at different orientations (Tramell & Goodrich 2002). Several bow like features are seen within the lobes as well as two bow-shaped structures at the tip of the lobes. Tramell & Goodrich interpreted these collimated lobes as three highly collimated outflows emanating from the central source of CRL 618. This interpretation is supported by the presence of multiple, high-velocity, molecular outflows aligned with the optical jets detected by Cox et al. (2003). The brightest emission occurs near the tip of each of the lobes, and there is no forbidden emission line seen in the central regions (Tramell & Goodrich 2002). These bow-shaped structures move outwards at ∼ 80 km s$^{-1}$ (S´anchez Contreras et al. 2002).

Several authors have derived the shock velocities from emission line ratios of the outer bow shock structures. Riera, Phillips & Mampaso (1990) concluded that two shock components were present (a low velocity shock with V$_s$ ∼ 20 km s$^{-1}$, and a faster shock with V$_s$ ∼ 90 km s$^{-1}$). TDG concluded that CRL 618 is a shock-excited N-rich object (with V$_s$ from 40 to 60 km s$^{-1}$), and S´anchez Contreras et al. (2002) derived shock velocities in the range from 75 to 200 km s$^{-1}$ as obtained from diagnostic line ratios or emission line profiles. The shock velocity derived by S´anchez Contreras et al. (2002) from the comparison of the observed emission line ratios with the models of HRH is slightly larger (but still comparable) than other estimates as a result of the higher [O III] flux measured by them. The shock velocity derived by S´anchez Contreras et al. (2002) from the emission line profiles is larger than that derived from the observed emission line ratios. As already discussed by S´anchez Contreras et al., similar differences have been observed in HH objects.

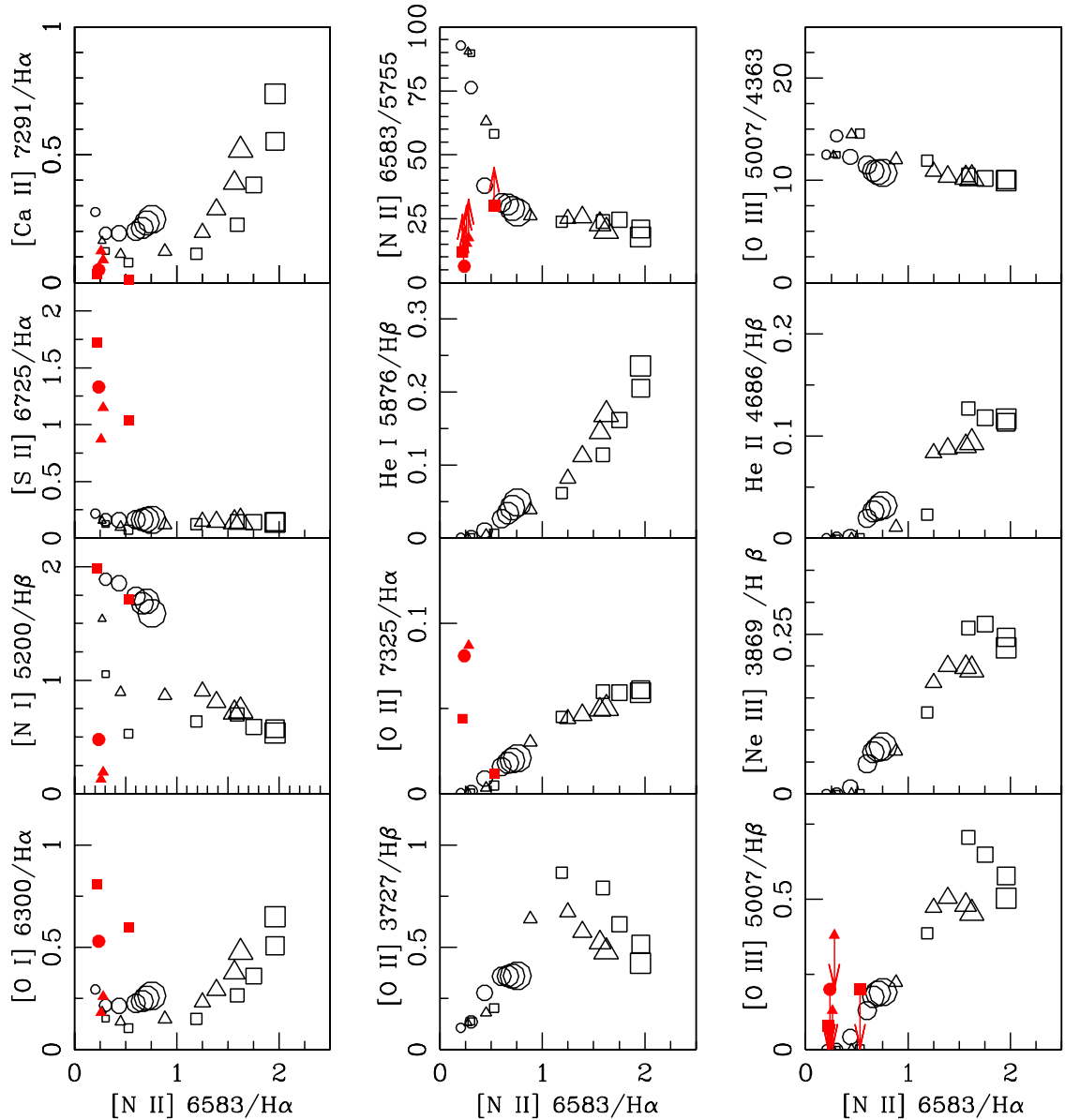

**Fig. 7.** As figure 5 for Type I PNe abundances and "local equilibrium preionization" conditions. The observed emisison line ratios of M 1-92 (red triangles), M 2-56 (red circles) and CRL 618 (red squares) are also shown in these plots.

Recently Lee & Sahai (2003) presented hydrodynamical simulations of a collimated fast wind interacting with a spherical AGB wind in order to reproduce the morphology and kinematics of CRL 618. They reproduced the presence of a bow-like feature at the tip of the lobes of CRL 618, where emission line fluxes peak. Lee & Sahai (2003) predicted absolute [S II] flux and several emission line ratios at the tip of the lobes in agreement within a factor of 2 with the observed values.

The emission line ratios of CRL 618 (from TDG) are plotted in Fig. 7, where we also included the predictions of the "local equilibrium preionization" mean Type I PNe abundance bow shock models. The high [N I] 5200 /Hβ emission line ratios observed at the tips of the lobes of CRL 618 strongly suggest that the nebula is nitrogen enriched. The position of these observations in the [N I]/Hβ vs. [N II]/Hα plot indicates that the emission may be formed in a low velocity bow-shock (favouring p=2). The [O III]/Hβ emission line ratios of CRL 618, which are upper limits, are also consistent with a low velocity bow-shock. However, other diagnostic line ratios are not reproduced by these models. The observed [O I]/H$\alpha$ and [S II]/H$\alpha$ are higher than the values expected for a low velocity bow shock, while the observed [Ca II]/H$\alpha$ line ratios are much lower than the predicted values. The problem with the [N II] 6583/5755 line ratios (i.e. the observed values are considerably lower than the predictions of the bow shock models) is also present in CRL 618.

At the low bow-shock velocity required to reproduce the spectra of CRL 618, the predicted [N II] 6583 / 5755 will decrease if the preshock density increases as shown by HRH. The electron densities measured in the lobes of CRL 618 by Sánchez Contreras et al. (2002) are 5000 to $10^4$ cm$^{-3}$, implying a pre-shock density of few $10^3$ cm$^{-3}$. An increase in the preshock density of the bow shock models would contribute to a moderate increase of the [O I] 6300/H$\alpha$ line ratios as required to fit the observed ratios. However, it would produce a decrease of the [N I] 5200/H$\beta$, [O II] 3727/H$\beta$ and [S II] 6725/H$\alpha$ ratios.

We conclude that CRL 618 is a nitrogen enriched nebula, with a N/H abundance of the order of the mean value for Type I PNe. The emission arising from the tips of the lobes can be roughly reproduced by a bow shock model with a velocity of about 50 - 75 km s$^{-1}$ assuming the "local equilibrium preionization" condition.

## 7. Conclusions

We have presented new HST STIS spectroscopy of Hen 3-1475, which allows us to spatially resolve the emitting knots, and to determine the excitation conditions of the innermost and intermediate knots from their optical spectra.

We have presented a simple model to predict the emission spectra of a bow shock for shocked features observed in proto-planetary nebulae (with a number of limitations). We present the effects of the pre-ionization conditions, the bow shock velocity, the bow shock shape and the gas abundances. We have shown that, apart from the bow shock velocity and gas abundances, the preionization conditions are a key factor in determining the location of the model predictions on the diagnostic diagrams.

We should keep in mind that the comparisons between the observed and predicted (from 3/2-D bow shock models) line ratios is quite uncertain. This is, however, a general problem found in comparisons between observed and predicted shock excited spectra (see, e.g. Raga, Böhm & Cantó 1996) as a result of the fact that the ionization/excitation in shocks is extremely sensitive to the local pre-shock conditions, which change both with position and time.

The spectra of the knots of Hen 3-1475 can be marginally reproduced by these simple bow shock models, for bow shock velocities of $\sim$ 150 to 200 km s$^{-1}$ assuming the "full preionization" condition (i.e. an uniform ionization, the ionization fractions of H$^+$ and He$^{++}$ being set to 1) within the observational uncertainties and the modeling approximations. The brightest and highest excitation knot NW1a is not well reproduced by the bow shock models presented here. This issue together with the fact that NW1a is the only knot detected in X-ray emission strongly suggest a bow shock velocity larger than the values needed to explain the shocked emission of the SE1 subcondensations (see Section 6.1). We note that the differences between the emission line profiles of the NW and SE knots also suggest that both features are geometrically and kinematically different (see figure 5 of Riera et al. 2003).

We have also applied this analysis to three proto-planetary nebulae (M 1-92, M 2-56 and CRL 618) with emitting features that resemble bow shock structures. To fit the observed spectra, the bow shock velocities required are $\leq$ 75 km s$^{-1}$ for M 2-56 and CRL 618, while M 1-92 requires larger bow shock velocities (from $\leq$ 100 to 150 km s$^{-1}$). In all cases the best fit corresponds to the "local equilibrium preionization" models.

From the comparison between the observed and predicted emission line ratios of four PPNe we have found that the poorest agreement with the observations is found for the [S II]/H$\alpha$, [O I]/H$\alpha$ and [N II] 6583/5755 emission line ratios. The observed [S II]/H$\alpha$ and [O I]/H$\alpha$ are well above the predicted values in all nebulae, while the observed [N II] 6583/5755 emission line ratios are lower than the predicted values.

*Acknowledgements.* The work of AR was supported by a grant from the Spanish Ministry of Science and Technology (AYA 2022-00205). LB acknowledges finantial support from CONACyT grant 40096-E and the UNAM PAPIIT grants 113002 and 118905.